\documentclass[prl,twocolumn,showpacs,preprintnumbers,amsmath,amssymb]{revtex4-1}

\usepackage{epsfig}
\usepackage{graphicx}
\usepackage{dcolumn}
\usepackage{bm}

\newcommand{\EQ}{\begin{equation}}
\newcommand{\EN}{\end{equation}}
\newcommand{\ea}{\end{eqnarray}}
\newcommand{\ba}{\begin{eqnarray}}

\newcommand{\bear}{\begin{eqnarray}}
\newcommand{\ear}{\end{eqnarray}}

\begin{document}
\title{Temperature dependence of charge transport in the half-filled 1D Hubbard model}


\author{J. M. P. Carmelo}
\affiliation{Center of Physics of University of Minho and University of Porto, LaPMET, P-4169-007 Oporto, Portugal}
\affiliation{CeFEMA, Instituto Superior T\'ecnico, Universidade de Lisboa, LaPMET, Av. Rovisco Pais, P-1049-001 Lisboa, Portugal}
\author{P. D. Sacramento}
\affiliation{CeFEMA, Instituto Superior T\'ecnico, Universidade de Lisboa, LaPMET, Av. Rovisco Pais, P--1049-001 Lisboa, Portugal}



\begin{abstract}
The use of hydrodynamic transport theory seems to indicate that the charge diffusion constant $D$ of the one-dimensional (1D) half-filled Hubbard 
model, whose Drude weight vanishes, diverges for temperature $T>0$, which would imply anomalous superdiffusive charge transport. Here the 
leading term of that constant is derived for low finite temperatures $k_B T/\Delta_{\eta}\ll 1$
where $\Delta_{\eta}$ is the Mott-Hubbard gap. It only diverges in the $T\rightarrow 0$ limit, being finite and decreasing upon 
increasing $T$ within the $k_B T/\Delta_{\eta}\ll 1$ regime. Our exact results both provide valuable physical information on
a complex quantum problem and bring about the interesting unsolved issue of how charge transport evolves from normal 
diffusive for $k_B T/\Delta_{\eta}\ll 1$ to anomalous superdiffusive in the $k_B T\rightarrow\infty$ limit.
\end{abstract}

\maketitle

The repulsive one-dimensional (1D) Hubbard model \cite{Lieb_68,Takahashi_72,Martins_98} is the paradigmatic quantum 
system for low-dimensional strongly correlated electron systems and materials. The real part of the charge conductivity
of that model reads,
\begin{equation}
\sigma (\omega,T) = 2\pi D^z (T)\delta (\omega) + \sigma_{\rm reg} (\omega,T) \, .
\label{sigma}
\end{equation}
When the charge stiffness $D^z (T)$ in its singular part is finite, the dominant charge transport is ballistic.
It is well established that in the thermodynamic limit the charge stiffness
exactly vanishes at half filling for all temperatures $T>0$ \cite{Ilievski_17,Carmelo_18}. 

The use of hydrodynamic theory and Kardar-Parisi-Zhang (KPZ) scaling
to study dynamical scaling properties of charge transport associated with the regular
part of the conductivity $\sigma_{\rm reg} (\omega,T)$
seems to indicate that the diffusion constant $D$ diverges for $T>0$, which would imply 
anomalous superdiffusive charge transport \cite{Ilievski_18,Fava_20,Moca_23}. However, the charge transport associated 
with the regular part of dc charge conductivity that characterizes the sub-ballistic timescales is a very complex quantum problem. 

In this Letter we derive the exact expression of the leading term of the charge diffusion constant 
$D$ of the half-filled 1D Hubbard model for magnetic fields $h\in [0,h_c]$
in the $k_B T/\Delta_{\eta}\ll 1$ regime. Here $\Delta_{\eta}$ is the Mott-Hubbard gap
and $h_c$ is the critical magnetic field for $T=0$ fully polarized ferromagnetism. That constant is found 
to diverge only in the $T\rightarrow 0$ limit, being finite and decreasing upon increasing $T$ within the $k_B T/\Delta_{\eta}\ll 1$ regime. 
This reveals normal diffusive charge transport in that low-temperature regime.

Our exact results thus contradict the expectation of Refs. \onlinecite{Ilievski_18,Fava_20,Moca_23} that the anomalous superdiffusive charge
transport found for hight temperatures $T\rightarrow\infty$ applies to all finite temperatures $T>0$. The regular part of dc 
charge conductivity is thus a quantum problem more involved than expected that deserves further investigations. 
In addition to provide valuable physical information on that complex quantum problem, our results bring about 
the interesting unsolved issue of how charge transport evolves from normal diffusive for $k_B T/\Delta_{\eta}\ll 1$ to anomalous 
superdiffusive for $k_B T\rightarrow\infty$.

The 1D Hubbard model describes $N$ electrons in a lattice of length $L$ with $N_a$ sites.
We consider the thermodynamic limit, its Hamiltonian at chemical potential $\mu=0$ in a magnetic field magnetic field $h\in [0,h_c]$ 
and under periodic boundary conditions reading,
\begin{equation}
\hat{H} = -t\sum_{\sigma, j}\left[c_{j,\sigma}^{\dag}\,c_{j+1,\sigma} + 
{\rm h.c.}\right] + U\sum_{j}\hat{\rho}_{j,\uparrow}\hat{\rho}_{j,\downarrow} + \mu_0 h {\hat{S}}_{s}^{z} \, .
\label{H}
\end{equation}
Here $c_{j,\sigma}^{\dag}$ creates one electron of spin projection $\sigma$ at site $j$,
$\hat{\rho}_{j,\sigma}= (\hat{n}_{j,\sigma}-1/2)$, $\hat{n}_{j,\sigma}=c_{j,\sigma}^{\dag}\,c_{j,\sigma}$,
${\hat{S}}_{s}^{z}={1\over 2}\sum_{j=1}^{L}(\hat{n}_{j,\uparrow} - \hat{n}_{j,\downarrow})$,
$\mu_0 = 2\mu_B$, and $h_c = (\sqrt{(4t)^2 + U^2} - U)/\mu_0$.

At fixed field $h<h_c$ and $T=0$, fully polarized ferromagnetism is also achieved upon increasing $U$ above 
$U_c = {8t^2\over \mu_0 h} -{1\over 2}\mu_0 h$. Upon increasing $h$, $U_c$ continuously decreases from 
$U_c = \infty$ at $h=0$ to $U_c = 0$ at $h = h_c$. The densities read
$m_{\eta}^z = (N_a-N)/L$ and $m_{s}^z = (N_{\uparrow} - N_{\downarrow})/L$.
We use natural units in which the Planck constant, electronic charge, and lattice spacing
are equal to one, so that $N_a=L$.

In the last three to four decades, representations in terms of spinons and holons have been widely used to successfully describe the 
static and dynamical properties of both integrable 1D models and the physics of the materials they describe \cite{Pereira_12}. 
Hence such representations became the paradigm of the 1D physics. Although the quantum
problem under study here is found to be defined in a small subspace for $k_B T/\Delta_{\eta}\ll 1$, the suitable identification 
of the charge carriers requires the use of a general representation that refers to the whole 
Hilbert space beyond the holon-spinon paradigm \cite{Carmelo_18}. 

For $u=U/4t>0$ the global symmetry of the 1D Hubbard model at chemical potential $\mu =0$ and magnetic field $h=0$
is larger than $SO (4) = [SU (2)\otimes SU(2)]/Z_2$, as usually assumed, and given by $[SO (4)\otimes U(1)]/Z_2$, which can be written as  
$[SU (2)\otimes SU(2)\otimes U(1)]/Z_2^2$ and $SO (3) \otimes SO (3) \otimes U (1)$ \cite{Carmelo_10}. 
This is consistent with the form of the two monodromy matrices of the Bethe-ansatz solution inverse-scattering 
method \cite{Martins_98,SM,Faddeev_81}. The corresponding $\eta$-spin $S_{\eta}$, spin $S_s$, $S_{\eta}^z= {1\over 2}(L-N)$, 
and $S_s^z ={1\over 2}(N_{\uparrow} - N_{\downarrow})$ are good quantum numbers. Singlet and multiplet refer in the following 
to $\eta$-spin or spin configurations.

Within the representation used here \cite{Carmelo_18}, the Bethe-ansatz solution
performs a squeezed-space construction \cite {Kruis_04}. It leads to (i) a set of $n=1,...,\infty$ $\eta n$-squeezed effective lattices, 
(ii) a set of $n=1,...,\infty$ $s n$-squeezed effective lattices, and (iii) a $\tau$-squeezed effective lattice whose occupancy configurations 
generate irreducible representations of the (i) $\eta$-spin $SU (2)$ symmetry, (ii) spin $SU (2)$ symmetry, and (iii) 
$\tau$-translational $U(1)$ symmetry, respectively, in $[SU (2)\otimes SU(2)\otimes U(1)]/Z_2^2$. 
Each such squeezed effective lattice refers to a corresponding
$\eta n$-band, $s n$-band, and $\tau$-band, respectively, whose discrete momentum values $q_j$ such that $q_{j+1}-q_j = 2\pi/L$ are 
Bethe-ansatz quantum numbers multiplied by $2\pi/L$ \cite{Carmelo_18}. 

The squeezed-space construction is such that the (a) $N_{\tau}$ $\tau$-particles and (b) 
$N_{\tau}^h = L - N_{\tau}$ $\tau$-holes of the $\tau$-squeezed effective lattice 
refer for $u>0$ to (a) singly occupied sites and (b) unoccupied and doubly occupied sites, respectively, of the electrons
for $u\rightarrow\infty$. The corresponding (a) spin and (b) $\eta$-spin degrees of freedom are described by the
occupancies of (a) $L_s = N_{\tau}$ physical spins $1/2$ and (b) $L_{\eta} = N_{\tau}^h$ physical $\eta$-spins $1/2$ in the
(a) $s n$-squeezed effective lattices and (b) $\eta n$-squeezed effective lattices, respectively. (A $u>0$ physical $\eta$-spin of
projection $+1/2$ and $-1/2$ refers to a $u\rightarrow\infty$ unoccupied and a doubly occupied site by electrons, respectively.)

The two $SU (2)$ symmetries in $[SU (2)\otimes SU(2)\otimes U(1)]/Z_2^2$ impose that 
for {\it all} energy eigenstates the number $L_{\alpha}$ can be written as
$L_{\alpha} = N_{\alpha} + {\cal{N}}_{\alpha}$. 
Here $N_{\alpha} = N_{\alpha,+1/2}+N_{\alpha,-1/2} = 2S_{\alpha}$ 
where $N_{\alpha,\pm 1/2}= S_{\alpha} \pm S_{\alpha}^z$ is the number of 
{\it unpaired physical spins} ($\alpha=s$) and {\it unpaired physical $\eta$-spins} ($\alpha=\eta$) 
that participate in a multiplet configuration and
${\cal{N}}_{\alpha}  = {\cal{N}}_{\alpha,+1/2} + {\cal{N}}_{\alpha,-1/2}$ 
where ${\cal{N}}_{\alpha,\pm 1/2} = L_{\alpha}/2 - S_{\alpha}$
is that of {\it paired physical spins} ($\alpha=s$) and {\it paired physical $\eta$-spins} ($\alpha=\eta$) that
participate in singlet configurations. All the latter are contained in $\alpha n$-pairs, each
involving a number $n=1,...,\infty$ of unbound ($n=1$) and bound $(n>1)$ singlet pairs.
The number $N_{\alpha n}$ of such pairs thus obeys the sum-rule 
$\sum_{n=1}^{\infty}2n\,N_{\alpha n} = {\cal{N}}_{\alpha}$. The $N_{\alpha n}$
$\alpha n$-pairs occupy  a number $N_{\alpha n}$ of sites of the corresponding $\alpha n$-squeezed effective lattice.
The corresponding $N_{\alpha n}^h = N_{\alpha} + \sum_{n'=n+1}^{\infty}2(n'-n)\,N_{\alpha n'}$
holes of that lattice include the $N_{\alpha}$ unpaired physical spins ($\alpha=s$) or unpaired physical $\eta$-spins ($\alpha=\eta$)
and a subset of $\sum_{n'=n+1}^{\infty}2(n'-n)\,N_{\alpha n'}<{\cal{N}}_{\alpha}$ paired physical spins ($\alpha=s$) or unpaired physical 
$\eta$-spins ($\alpha=\eta$) from $n'>n$ $\alpha n'$-branches. 

The Hamiltonian, Eq. (\ref{H}), in the presence of a uniform vector potential (twisted boundary conditions) remains solvable 
by the Bethe ansatz \cite{Carmelo_18}. The energy eigenstates described by the Bethe ansatz are $\eta$-spin highest-weight 
states (HWSs) \cite{SM} whose $N_{\eta} = 2S_{\eta}$ unpaired physical $\eta$-spins have the same projection 
$+1/2$. For them, the momentum eigenvalues, $P (\Phi)$, have the general form,
\begin{equation}
P (\Phi/L) = P (0) - (L_{\eta}-\sum_{n}2n\,N_{\eta n}){\Phi\over L} =  P (0) - N_{\eta}\,{\Phi\over L} \, ,
\label{PeffU}
\end{equation}
where $P (0)$ is their value for $\Phi = 0$ \cite{Carmelo_18}.
The term $- L_{\eta}\,{\Phi\over L}$ in $-(L_{\eta}-\sum_{n}2n\,N_{\eta n}){\Phi\over L}$
refers to {\it all} $L_{\eta}$ physical $\eta$-spins coupling to the vector potential
in the absence of physical $\eta$-spins singlet pairing. Indeed, the coupling counter terms $\sum_{n}2n\,N_{\eta n}{\Phi\over L}$ 
refer to the number $2n$ of paired physical $\eta$-spins in each $\eta n$-pair. 
These counter terms {\it exactly cancel} the coupling of the 
corresponding $2n$ paired physical $\eta$-spins in each such an $\eta n$-pair. 
As a result of such counter terms, only the $N_{\eta} = L_{\eta}-\sum_{n}2n\,N_{\eta n}$
unpaired physical $\eta$-spins couple to the vector potential and thus carry charge current.
For non-HWSs, the result is $P (\Phi/L) = P (0) - (N_{\eta,+1/2}-N_{\eta,-1/2})\,{\Phi\over L}$.

Hence for {\it all} energy eigenstates the {\it charge carriers} are the unpaired physical $\eta$-spins. 
One can generate from each $\eta$-spin HWS for which $N_{\eta} = N_{\eta,+1/2}$ a set
of $n_{\eta}^z = 1,...,2S_{\eta}$ non-HWSs populated by $N_{\eta,+1/2} = 2S_{\eta}-n_{\eta}^z$
and $N_{\eta,-1/2} = n_{\eta}^z$ unpaired physical $\eta$-spins with opposite projection
and thus opposite coupling to the vector potential \cite{SM}. The elementary charge currents $j_{\pm 1/2}$ carried
by each unpaired physical $\eta$-spin that populates a HWS and its non-HWSs 
is then found to read \cite{Carmelo_18},
\begin{equation}
j_{\pm 1/2} = \pm {\langle\hat{J}_{HWS}^z (S_{\eta})\rangle\over 2S_{\eta}} = \pm {\langle\hat{J}_{HWS}^z (S_{\eta})\rangle\over N_{\eta}} \, .
\label{J-eta-spin-spin}
\end{equation}
Here $\langle\hat{J}_{HWS}^z (S_{\eta})\rangle$ denotes the charge current expectation
value of the HWS. That of its non-HWSs reads $\langle\hat{J}^z  (S_{\eta}^z)\rangle = 
\sum_{\sigma =\pm 1/2} j_{\sigma}\,N_{\eta,\sigma}$, which can be written as
$\langle\hat{J}^z  (S_{\eta}^z)\rangle = {S_{\eta}^z\over S_{\eta}}\langle\hat{J}_{HWS}^z (S_{\eta})\rangle$.
This applies to {\it all} finite-$S_{\eta}$ energy eigenstates. 

The $\mu =0$ ground states and all other $S_{\eta}=0$ energy eigenstates have no charge carriers, so that their 
charge current expectation value vanishes. For general finite-$S_{\eta}$ energy eigenstates, a number 
$N_{\eta} = 2S_{\eta}$ of holes out of both the (i) $N_{\tau}^h = N_{\eta} + {\cal{N}}_{\eta}$
$\tau$-band holes and (ii) $N_{\eta n}^h = N_{\eta} + \sum_{n'=n+1}^{\infty}2(n'-n)\,N_{\eta n'}$ holes 
of $\eta n$-bands with finite occupancy, $N_{\eta n}>0$,
describe the translational degrees of freedom of the $N_{\eta}$ unpaired physical $\eta$-spins that carry the
elementary charge currents $j_{\pm 1/2}$, Eq. (\ref{J-eta-spin-spin}). 

However, ${\cal{N}}_{\eta}=0$ and thus $N_{\eta n}=0$ for $n=1,...,\infty$ for the states that contribute 
to the quantum problem studied in the following. The
translational degrees of freedom of their unpaired physical $\eta$-spins are thus described only 
by the $N_{\tau}^h=N_{\eta}$ $\tau$-holes. In addition, $N_{s n}=0$ for $n>1$, their $s1$-band 
being full for $s1$-band momentum values $q \in [-k_{F\downarrow},k_{F\downarrow}]$ and empty for
$\vert q\vert \in [k_{F\downarrow},k_{F\uparrow}]$ where $k_{F\uparrow} = {\pi\over 2}(1+ m_s^z)$
and $k_{F\downarrow} = {\pi\over 2}(1- m_s^z)$ \cite{SM}.

The charge diffusion constant $D$ can be derived by accounting for the Kubo linear response formula for 
the real part of dc charge conductivity being related to it through the Einstein relation. 
At very low temperatures, the charge currents contributing to it 
are generated by processes where upon moving in the $\tau$-squeezed effective lattice the $N_{\tau}^h=2S_{\eta}$
$\tau$-holes that describe the translational degrees of freedom of the $N_{\eta}=2S_{\eta}$ 
unpaired physical $\eta$-spins interchange position with the 
$N_{\tau}=L-2S_{\eta}$ $\tau$-particles. In the following we identify the motion of the $N_{\tau}^h=2S_{\eta}$
holes with that of such $N_{\eta}=2S_{\eta}$ $\eta$-spin-$1/2$ charge carriers they translationally describe. 

The physics associated with the diffusive charge transport 
is for $U>0$ and $\mu =0$ such that in the limit of very low temperatures nearly ballistic transport 
occurs, but with zero charge stiffness. Nearly ballistic transport with zero spin stiffness
also occurs in the gapped spin-$1/2$ $XXZ$ chain \cite{Karrasch_14}. In the present case,
the corresponding charge carriers are actually $\eta$-spin triplet pairs that in the 
$\tau$-squeezed effective lattice refer to two adjacent unpaired physical $\eta$-spins with the same projection $+1/2$
or $-1/2$. We call them {\it $\eta$-spin triplet charge carriers} to distinguish from their two unpaired physical $\eta$-spins. 

Relative to the $\mu=0$ ground states,
the excitation energy and momentum of one such a $S_{\eta}^z=\pm 1$ $\eta$-spin triplet pair read,
\begin{equation}
\delta E = - \varepsilon_{\tau} (q) - \varepsilon_{\tau} (q') + 2\mu_B h \hspace{0.20cm}{\rm and}\hspace{0.20cm}
k \equiv \delta P  = \pm \pi - q - q' \, ,
\label{deltaE}
\end{equation}
respectively, where the $\tau$-band energy dispersion $\varepsilon_{\tau} (q)$ is defined in Ref. \onlinecite{SM}
for $h \in [0,h_c]$. Here $-\varepsilon_{\tau} (q)>0$ and $-\varepsilon_{\tau} (q')>0$ are the energies for
creation of two corresponding $\tau$-holes of $\tau$-band momentum $-q$ and $-q'$, respectively,
and the $\tau$-band momentum values shift $\pm \pi/L$ that gives the $k$'s term $\pm\pi$ and 
the energy $2\mu_B h$ are due to a $s1$-band occupancy deviation $\delta N_{s1}=-1$ \cite{SM}.

At very low temperature the excitation energy of such $\eta$-spin triplet 
pairs, Eq. (\ref{deltaE}), must be very near its minimum allowed value, the Mott-Hubbard gap,
\begin{equation}
\Delta _{\eta} = \Delta _{\tau} + 2\mu_B h \, ,
\label{gap}
\end{equation}
where $\Delta _{\tau} = - 2\varepsilon_{\tau} (\pi)$.
It is plotted in Fig. 1 (a) of Ref. \onlinecite{SM}, its general expression for $h\in [0,h_c]$ being given in 
that reference. It involves the excitation energy of the two $\tau$-holes, Eq. (\ref{deltaE}), at
$-q = \pm \pi$ and $-q' = \pm \pi \mp 2\pi/L$. 

The translational degrees of freedom of the $\eta$-spin triplet charge carriers that play the role of low-temperature 
diffusing charges are thus described by two $\tau$-holes whose $\tau$-band momentum values $-q$ and $-q'$
are very near $-\pi$ or $\pi$, $-q \approx \mp\pi$ and $- q' = - q \pm 2\pi/L$. This gives
an excitation momentum $k =  \pm \pi - 2q \pm 2\pi/L$ that reads
$k = \pm \pi - 2q \approx \mp \pi$ in the thermodynamic limit, the corresponding two unpaired physical $\eta$-spins with the same 
projection $+1/2$ or $-1/2$ being adjacent in the $\tau$-squeezed effective lattice upon moving with nearly the 
same $\tau$-band momentum.

$S_{\eta}^z=0$ $\eta$-spin triplet pairs have the same minimal excitation energy $\Delta _{\eta}$ yet involve two $\eta$-spin carriers with 
{\it opposite} projection. They do not not contribute to charge transport because their elementary 
charge currents, Eq. (\ref{J-eta-spin-spin}), cancel each other. The same applies to {\it singlet} $\eta 1$-pairs, which 
are the only $\eta n$-pairs whose minimal excitation energy is also $\Delta _{\eta}$, Eq. (\ref{gap}).

\begin{figure*}
\includegraphics[width=0.495\textwidth]{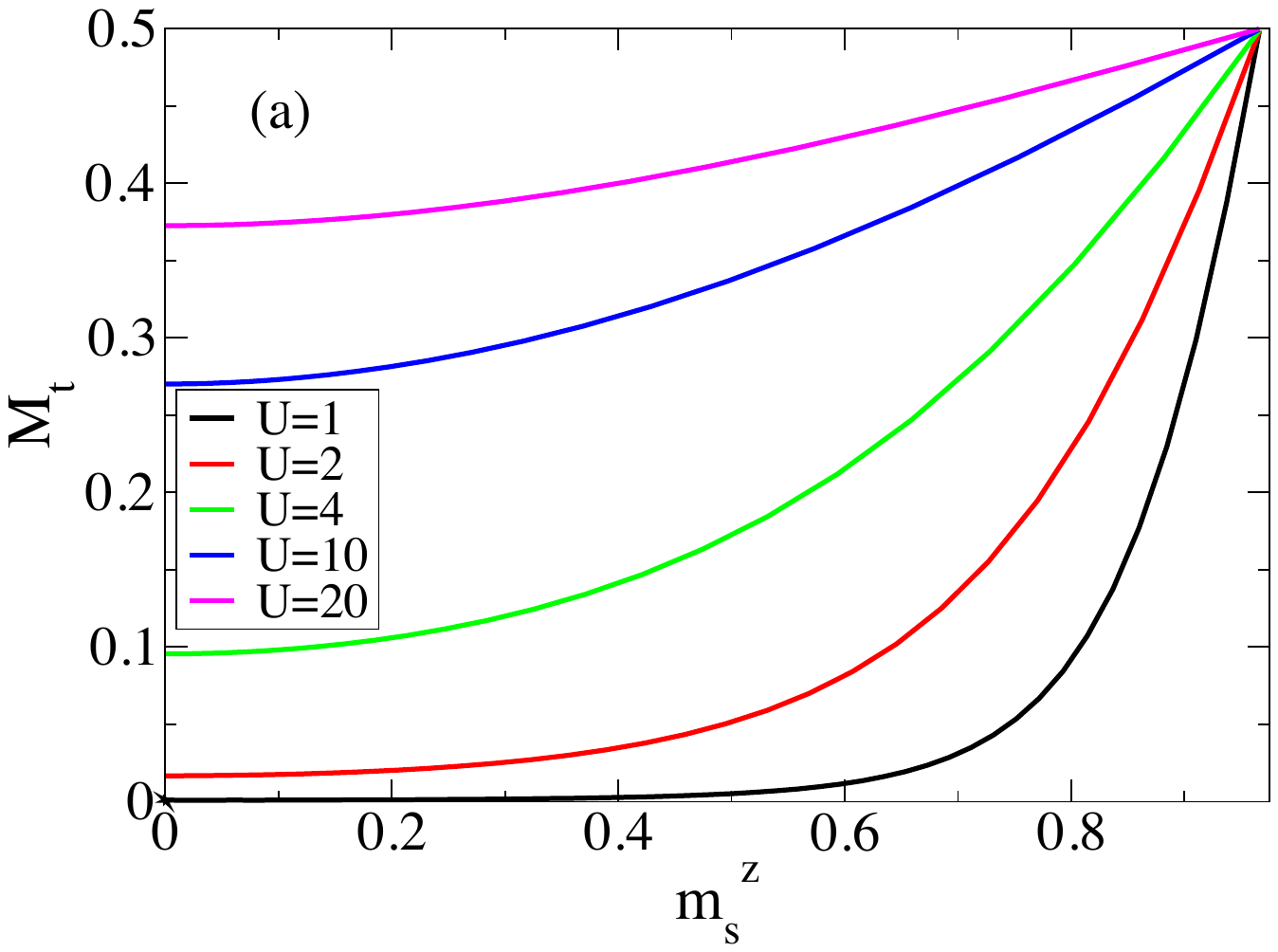}
\includegraphics[width=0.495\textwidth]{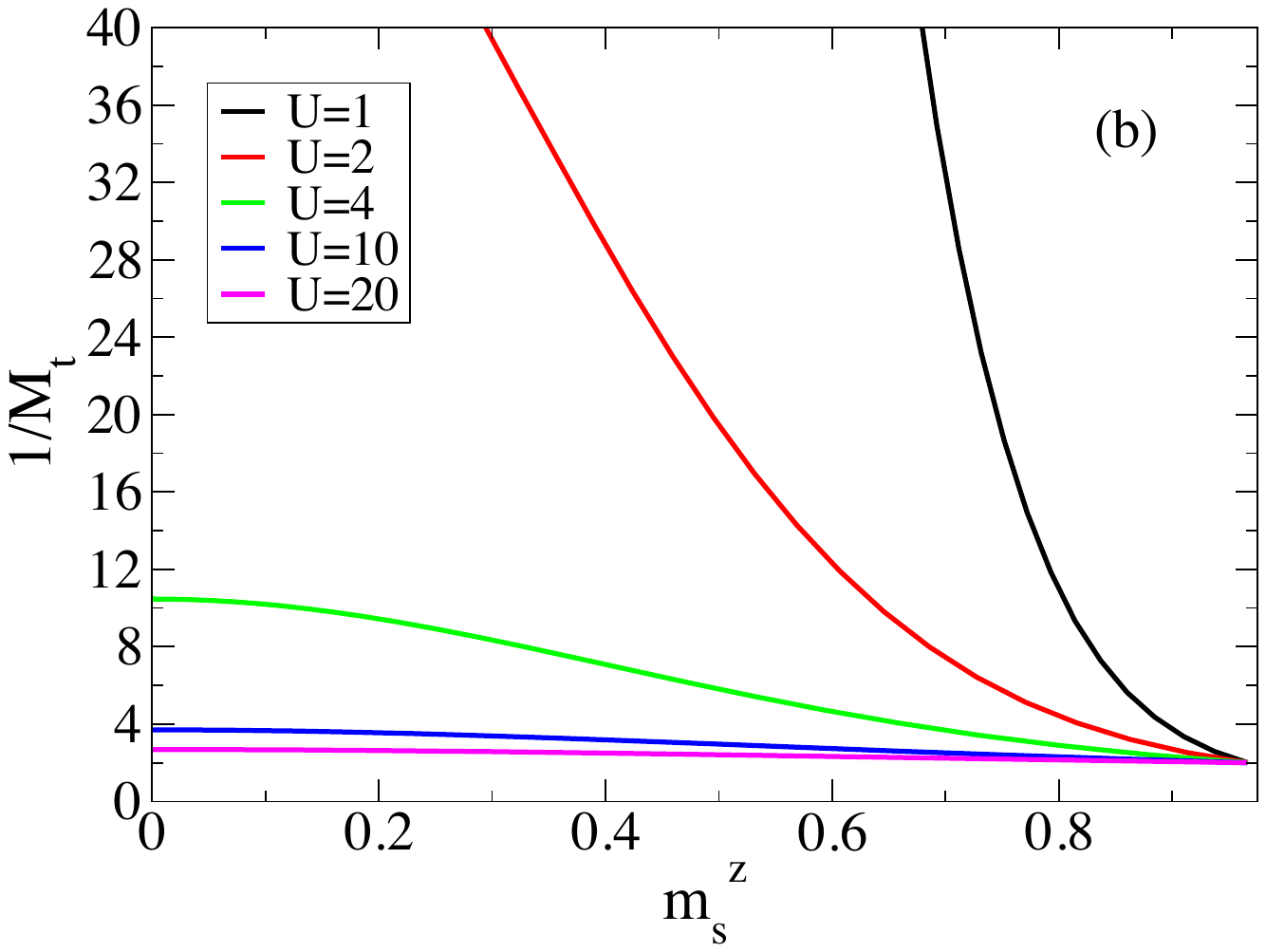}
\caption{The charge transport mass $M_t=M_{\eta}/2$ of the $\eta$-spin triplet pairs 
appearing in Eqs. (\ref{jqq}) and (\ref{D}) (a) and its inverse $1/M_t$ (b) plotted for 
several values of $U$ in units of transfer integral $t$ as a function of spin density for the interval
$m_z^z = 2S_s^z/L\in [0,1]$ that refers to magnetic fields $h \in [0,h_c]$. The charge transport
mass $M_t =M_{\eta}/2 \in [0,1/2t]$ vanishes in the $U\rightarrow 0$ limit, consistently with charge ballistic 
transport at $U=0$, and is enhanced upon increasing $u=U/4t$ and/or the magnetic field $h$.}
\label{figure1PRL}
\end{figure*}

The lowest excited states have a single $\tau$-squeezed effective lattice domain wall that refers to
a single $\eta$-spin triplet pair. The domain wall or diffusing charge associated with the $\eta$-spin
triplet pair moves ballistically for such lowest excited states. Also for excited states with a finite number and thus vanishing 
concentration of $\eta$-spin triplet pairs, such pairs move ballistically over large $\tau$-squeezed effective lattice
distances before interacting with other $\eta$-spin triplet pairs. They carry an elementary charge current
$j_{\pm 1} (k) = 2j_{\pm 1/2} (q)\vert_{q=(\pm\pi -k)/2}$ where $j_{\pm 1/2} (q)$ is that carried by each of the two corresponding adjacent 
unpaired physical $\eta$-spins, Eq. (\ref{J-eta-spin-spin}). It reads \cite{SM},
\begin{equation}
j_{\pm 1} (k) = \mp {(k \pm \pi)\over M_t}\hspace{0.20cm}{\rm for}\hspace{0.20cm}
k = \pm \pi - 2q \approx \mp \pi \, .
\label{jqq}
\end{equation}
Here $M_t$ is the charge transport mass of the $\eta$-spin triplet pairs such that
$1/M_t = \vert\partial j_{\pm 1} (k)/\partial dk\vert_{k=\mp\pi}$ whose
general expression for $h\in [0,h_c]$ is given in Ref. \onlinecite{SM}.
(See Fig. \ref{figure1PRL}.) 

Consistently with nearly ballistic charge transport, but with zero charge stiffness, 
the leading term of the charge diffusion constant $D$ in the
$k_B T/\Delta_{\eta} \ll 1$ regime can be expressed in terms of the first derivative with respect to $m_{\eta}^z$ 
of the $T=0$ charge stiffness $D^z$ for $m_{\eta}^z\ll 1$ \cite{SM}, 
\begin{equation}
D = {e^{\Delta_{\eta}\beta}\over 2}{d D^z\over d m_{\eta}^z}\vert_{m_{\eta}^z=0} = {e^{\Delta_{\eta}\beta} \over 4M_t}
\hspace{0.20cm}{\rm where}\hspace{0.20cm}D^z = {m_{\eta}^z\over 2M_t} \, .
\label{D}
\end{equation}
That $D \propto e^{\Delta_{\eta}\beta}$ for low temperature where $\beta = 1/k_B T$ applies
to gapped 1D models in the quantum Sine-Gordon universality class \cite{Damle_05}.
In the opposite limit of high temperature, there is for some models a relation between $D$ and the stiffness
that rather involves a second derivative \cite{Ilievski_18,Medenjak_17}.

To arrive to the expression, Eq. (\ref{D}), we treated the quantum problem as a dilute gas 
of excited $\eta$-spin triplet pairs, each with energy 
$\epsilon_{\eta} (k) = \Delta_{\eta} + {(k\pm \pi)^2\over 2M_{\eta}}$ and velocity 
$v_{\eta} (k) = {k\pm \pi\over M_{\eta}}$ where $M_{\eta}=2M_t$ such that $1/M_{\eta} = {d^2 \epsilon_{\eta} (k)\over d k^2}$
is their static mass \cite{SM}. Their motion and collisions dominate 
the charge transport properties, their spacing $x_{\eta} = (\beta /M_{\eta})^{1/2}\,e^{\Delta_{\eta}\beta}$
being for $k_B T/\Delta_{\eta} \ll 1$ much larger than their thermal de Broglie wavelength 
$\lambda_{\eta} = (\beta /M_{\eta})^{1/2}$. The quantum problem to be solved then refers to the collisions of the $\eta$-spin triplet pairs 
and corresponding $S$-matrix. However, the root mean square thermal velocity of such
pairs, $v_{\eta,\beta} = 1/\sqrt{M_{\eta}\beta}$, tends to zero as $k_B T/\Delta_{\eta} \rightarrow 0$,
so that the $S$-matrix that describes the present quantum problem refers to vanishing exchange incoming or outgoing 
momenta. 

This much simplifies the problem, the leading behaviors of the uniform charge 
susceptibility and real part of the dc charge conductivity being for the $k_B T/\Delta_{\eta} \ll 1$
regime and in our units expressed solely in terms of the two above length scales as $\chi  = (2/\pi)^{1/2}{\beta\over x_{\eta}}$ and 
$\sigma = (2/\pi)^{1/2}{\lambda_{\eta}\over 2}$, respectively \cite{SM,Carmelo_04}. The use of the Einstein relation, $\sigma = \chi D$, then 
gives the leading behavior $D = \sigma/\chi = {x_{\eta} \lambda_{\eta}\over 2\beta}$, so that
$D = e^{\Delta_{\eta}\beta}/(4M_t)$ for $k_B T/\Delta_{\eta}\ll 1$, as given in Eq. (\ref{D}).

The Mott-Hubbard gap $\Delta _{\eta}$, Eq. (\ref{gap}), in the expression, Eq. (\ref{D}), 
vanishes at $U=0$ and continuously increases upon increasing $u=U/4t$. It is plotted as a function 
of $u$ in Fig. 1 (a) of Refs. \cite{SM} and \cite{Carmelo_21} for $m_s^z=0$ ($h=0$) and $m_s^z =1$ $(h=h_c)$.
Interestingly, in spite of the apparently different expressions 
$\Delta _{\eta} = U - 4t + 8t\int_0^{\infty}d\omega {J_1 (\omega)\over\omega (1 + e^{2\omega u})}$
at $h=0$ where $J_1 (\omega)$ is a Bessel function and $\Delta _{\eta} = \sqrt{(4t)^2 + U^2} - 4t$ at $h=h_c$,
the two corresponding curves in the figure are very similar, which shows that charge-spin coupling effects
are very small at $\mu=0$. (The inset of that figure shows the amplified deviations between the two lines, 
the amplified general deviation $\Delta _{\eta} (m_s^z) - \Delta _{\eta} (0)$ being shown in Fig. 1 (b) 
of Ref. \onlinecite{SM} for $m_s^z\in [0,1]$.)

For fields $h \in [0,h_c]$, the inverse transport mass $1/M_t$ in Eqs. (\ref{jqq}) and (\ref{D})
decreases upon increasing $U$ from $\infty$ for $U\rightarrow 0$, 
consistently with the $U=0$ ballistic charge transport, to $2t$ at $U=U_c$. 
$M_t$ (a) and $1/M_t$ (b) are plotted in Fig. \ref{figure1PRL}
for several values of $U$ in units of $t$ as a function of spin density for the interval
$m_z^z = 2S_s^z/L\in [0,1]$ that refers to magnetic fields $h \in [0,h_c]$.

Concerning the physics behind our results, the present interesting case of 
nearly ballistic charge transport, but with zero charge stiffness, is
consistent with $D$ being proportional to the inverse transport mass of the $\eta$-spin triplet charge carriers.
Their concentration goes to zero for $k_B T/\Delta_{\eta}\rightarrow 0$ as $S_{\eta}/L = 1/x_{\eta} = (M_{\eta}/\beta)^{1/2}\,e^{-\Delta_{\eta}\beta}$. 
At the same time, the $\tau$-squeezed effective lattice 
average distance $x_{\eta} = (\beta /M_{\eta})^{1/2}\,e^{\Delta_{\eta}\beta}$ traveled by one $\eta$-spin 
triplet pair before it interacts with other such pairs and 
ceases to be ballistic diverges exponentially. The interplay 
between the exponential factors in the concentration of charge carriers and the distance between them
is behind $D$ diverging exponentially, Eq. (\ref{D}).
It becomes infinite in the $T\rightarrow 0$ limit, having large finite values at low finite temperatures.

In contrast to the results of Refs. \onlinecite{Ilievski_18,Fava_20,Moca_23}, which predict that the charge 
anomalous superdiffusive transport found for high temperatures applies to all finite temperatures $T>0$, we thus find that 
for $k_B T/\Delta_{\eta}\ll 1$ the charge diffusion constant leading term is finite and given by 
Eq. (\ref{D}), which implies normal diffusive charge transport. At very low temperatures within the $k_B T/\Delta_{\eta}\ll 1$ regime, 
the charge diffusion constant, Eq. (\ref{D}), actually decreases upon increasing the temperature. 

The anomalous superdiffusive spin transport at any finite temperature predicted by hydrodynamic theory (KPZ scaling) 
for the spin-$1/2$ $XXX$ chain \cite{SM,Ljubotina_19,Ljubotina_17}
and half-filled 1D Hubbard model is compatible with the gapless nature of spin excitations.
Its prediction of superdiffusive charge transport is then based on the assumption that at half filling $\eta$-spin and spin $SU(2)$ symmetries 
are related by duality, which would imply that spin and charge transport have identical properties.
However, the physical reason why concerning charge transport that prediction fails for $k_B T/\Delta_{\eta}\ll 1$ 
is that due to quantum effects the above duality relation is broken by the Mott-Hubbard gap \cite{SM}.

By both providing the exact expression of the charge diffusion constant $D$ in the $k_B T/\Delta_{\eta}\ll 1$ regime, 
Eq. (\ref{D}), and showing that finite-temperature charge transport is a much more complex quantum problem than predicted by
hydrodynamic theory, as at low temperature KPZ scaling is erased by {\it quantum} effects associated
with the Mott-Hubbard gap \cite{SM}, our results open a new avenue for the further understanding of the charge 
transport properties of the paradigmatic quantum system 
for low-dimensional strongly correlated electron systems, the 1D Hubbard model.

However, the results obtained by hydrodynamic theory and KPZ scaling in Refs. \onlinecite{Ilievski_18,Fava_20,Moca_23} 
concerning charge transport are valid at high temperatures $T\rightarrow\infty$ for which the above duality relation emerges, 
so that it is anomalous superdiffusive in that regime. Combining such a behavior with that found here for the $k_B T/\Delta_{\eta}\ll 1$ 
regime and accounting for the lack of phase transitions imposed by the Mermin-Wagner Theorem, 
there must be a crossover such that $D$ decreases upon enhancing $T$ until reaching a minimum at some intermediate
$U$-dependent temperature and then increases up to $D=\infty$ for $T\rightarrow\infty$. 
The full understanding of that crossover requires studies on charge transport in the 
half-filled Hubbard model for temperatures $T \approx \Delta_{\eta}/k_B$, which is
a very complex quantum problem beyond the goals of this Letter.

Our exact results apply to the (i) charge and (ii) spin degrees of freedom of the 1D Hubbard model for
(i) $\mu=0$ and $U>0$ and (ii) $h=0$ and $U<0$, respectively. They
provide valuable physical information on a very complex quantum problem \cite{SM,Tsvelik_87} and
show that finite-temperature charge (spin) transport in the $\mu=0$ repulsive ($h=0$
attractive) 1D Hubbard model deserves further studies to clarify the mechanisms under which it 
evolves from normal diffusive for $k_B T/\Delta_{\alpha}\ll 1$ to anomalous 
superdiffusive for $k_B T\rightarrow\infty$ where $\alpha =\eta$ ($\alpha =s$.)

\acknowledgements
We thank Toma\v{z} Prosen and Subir Sachdev for illuminating discussions.
J. M. P. C. and P. D. S. acknowledge support from Funda\c{c}\~ao para a Ci\^encia e Tecnologia
through the Grant UIDB/04650/2020 and 
UID/CTM/04540/2019, respectively.\\ \\ 

\end{document}